\documentstyle[aps,prl,floats,twocolumn,psfig]{revtex}
\def\cm{{\rm cm}}

\def\erg{{\rm erg}}

\def\eq{\begin{equation}}
\def\en{\end{equation}}
\def\eqa{\begin{eqnarray}}
\def\ena{\end{eqnarray}}

\begin{document}
\draft
\twocolumn[\hsize\textwidth\columnwidth\hsize\csname@twocolumnfalse%
\endcsname

\title{\bf How Sandcastles Fall}
\author{Thomas C. Halsey and Alex J. Levine}
\address{Exxon Research and Engineering, Route 22 East, Annandale,
N.J. 08801}
\date{\today}
\maketitle
\begin{abstract}
Capillary forces significantly affect the stability of sandpiles.  We
analyze the
stability of sandpiles with such forces, and find that the critical angle
is unchanged in the limit of an infinitely large system; however, this angle is
increased for finite-sized systems. The failure occurs in the bulk of the
sandpile
rather than at the surface.  This is related to a standard result in soil
mechanics.  The increase in the critical angle is determined by the surface
roughness of the particles, and exhibits three regimes as a function of the
added-fluid volume.  Our theory is in qualitative agreement with
the recent experimental results of Hornbaker {\em et al.}, although not
with the
interpretation they make of these results.
\end{abstract}

]

The continuum mechanics of most materials was established in the
$19$th century; however, the mechanics of granular
materials is still largely mysterious \cite{nagel}. The study of granular
media is
also motivated by the ubiquity of this form of matter in a variety of
industrial
contexts, as well as in geophysical ones.

While most recent attention has focussed on dry granular media, a
recent
experimental study by Hornbaker {\em et al.} has opened the relatively
unexplored
subject of ``humid" granular media, in which small amounts of added fluid
generate, through capillarity, adhesive forces between the grains
\cite{ndame}. Somewhat whimsically, these authors argue that their work is
appropriate for the understanding of sandcastles; we actually take this point
seriously, because adhesive forces and other liquid effects are extremely
important in geophysical applications, of which sandcastles are an unusual
example.

In this Letter, we present a theory of the stability of humid
sandpiles, based upon a continuum analysis of their statics. While previous
work
on the statics and dynamics of dry sandpiles concentrated on the behavior of
the pile's surface \cite{bouchaud}, we find that the addition of
small adhesive forces between the grains causes the site of failure to move
from the surface into the bulk of the sandpile, a fact well-known in soil
mechanics. Even though the failure of the sandpile at the critical angle
occurs at
some finite depth, in the limit of infinite system size, the critical angle is
actually unchanged by the adhesion.  For finite systems the angle of
repose is increased from the infinite-system/non-adhesive critical angle. By
analyzing the cohesive effect of small amounts of wetting fluid, we
find that this increase in the critical angle as a function of the added-fluid
volume exhibits a range over which the dependence is linear, in agreement
with the
principal result of Hornbaker {\em et al.} However, we disagree with the
suggestion
of these authors that most of the wetting fluid will be found outside the
particle contact zones.

To determine the stability of a sandpile, we must have a criterion for local
failure of the sandpile. For a non-adhesive (dry) sandpile, a simple
phenomenological criterion for failure is that
\eq
\tau > k \sigma,
\label{eq:dry}
\en
where $\tau$ is the tangential stress across some plane
interior to the sandpile, $\sigma$ is the normal compressive stress across that
plane, and
$k$ is the internal friction coefficient.  For a given stress state stability
requires that there be no plane for which the ratio of $\tau$ to $\sigma$
exceeds $k$.  To determine the plane on which this ratio is maximized we turn
to the Mohr construction \cite{ugaral}.

The Mohr circle provides a geometric construction to transform under rotations
a two dimensional symmetric tensor such as the stress tensor.  Suppose the two
principal stresses at a point are $\sigma_1$ and $\sigma_2$.  Then a circle
drawn in the $\sigma - \tau$ plane through the points $(\sigma_1,0) $ and
$(\sigma_2,0)$, symmetrically about the $\sigma$ axis, gives the normal and
tangential stresses on any plane.  To find these stresses for a
physical system rotated by an angle $\phi$ from the principal system one
looks at
the points on the circle at an angle $2 \phi$ from the horizontal.  These two
opposite points give the normal and tangential stresses across the physical
planes of the rotated system.  Thus the stability criterion for a dry
sandpile is
that no Mohr circle may extend above the line $\tau = k \sigma$.

Consider a semi-infinite dry sandpile, whose surface is oriented at an
angle $\theta$ to the horizontal. We choose an $x-z$ coordinate system, in
which $z$ gives the distance from the surface of the pile ($z > 0$ down), and
$x$ gives the distance parallel to the surface. Then the stress tensor
$\sigma_{ij}$ satisfies the static equations
\eqa
\partial_z\sigma_{zz} + \partial_x \sigma_{xz} & = & \rho g \cos \theta, \\
\partial_z\sigma_{zx} + \partial_x \sigma_{xx} & = & \rho g \sin \theta,
\ena
where $\rho$ is the density of the sandpile. To solve these equations, we first
restrict ourselves to solutions which are functions of $z$ alone--in a
semi-infinite system, any $x$-dependence of the solutions would be liable to
generate arbitrarily large stresses near the surface, which would cause the
system to buckle.   The most general $z$-dependent
solution, which also satisfies the boundary condition that the surface is
stress-free, is
\eqa
\label{eq:stress1}
\sigma_{zz} & = & \rho g z \cos \theta,\\
\sigma_{xz} & = & \rho g z \sin \theta,\\
\label{eq:stress3}
\sigma_{xx} & = & C (z),
\ena
The well-known stress indeterminacy in granular media  implies that
$C(z)$ is an unknown function. We can fix this function, however, by
finding the form of $C(z)$ that allows us to 
maximize the critical angle. The angle thereby obtained will certainly be an upper
bound on the true critical angle; furthermore, if the sandpile is free to adjust
its undetermined stress $C(z)$ within some range, we expect that this upper
bound will be identical to the critical angle. Thus, while  it
is impossible in general to fix $C(z)$ without some constitutive or closure relation, we do believe
that it is possible to fix it at the critical angle.

Clearly the function $C(z)$ that will maximize the critical angle will have the form
$C(z) = C^{\prime} z$.  To apply the
geometric Mohr construction, we first determine the diameter and the
position of
the center of the Mohr circle given by the difference and sum of
eigenvalues of the
stress tensor respectively.  From Eqs.\ (\ref{eq:stress1}-\ref{eq:stress3})
we find
that these eigenvalues are
\eq
\sigma_{1,2} = z \rho g \cos \theta \left( \frac{1 + c}{2} \pm
\sqrt{ \frac{(1+c)^2}{4} + \tan^2 \theta - c} \right),
\en
where $c = C^{\prime}/(\rho g \cos \theta) $.
Note that both the radius and center position of the Mohr circle depend
linearly on the depth $z$ below the surface.  The maximum value of
$\tau/\sigma$ as a function of $c$ is
\eq
\frac{\tau}{\sigma} \Big{\vert}_{max} = \sqrt{ \frac{(1+c)^2}{4(c-\tan^2
\theta)}-1}.
\label{eq:c}
\en
Now this quantity must be $< k$, so to find the maximum critical angle, we
wish to minimize the right-hand side of Eq.\ (\ref{eq:c}) with respect to $c$.
We then find that at this minimum value of $c$,
\eq
\frac{\tau}{\sigma} \Big{\vert}_{max} = \tan \theta,
\en
so that the critical angle is $\theta_c = \tan^{-1} k$, which is a classic
result \cite{class}.  This simple model, by
itself, does not indicate at which depth failure initiates--presumably a
dynamical model would resolve this ambiguity \cite{relax}. 

Now consider a sandpile in which a normal adhesive stress $s_A$ is exerted
across every plane, in addition to whatever other stresses may exist due to the
body forces. This stress introduces a normal force between pairs of contiguous
particles which allows the sandpile to support a finite shear
stress, even in the limit of zero applied compressive stress.  The maximum
supported shear stress, in this case, is $k s_A$ and we therefore replace
the dry
sand failure criterion, Eq.~(\ref{eq:dry}), by
\eq
\label{eq:wet}
\tau > k \left( \sigma + s_A \right) \mbox{     }.
\en

Performing a calculation similar to the one above, we find that the failure
criterion, which is now an explicit function of depth, is
\eq
\label{failure_criterion}
k = \tan \theta \big (1+ \frac{s_A}{\rho g z\cos \theta} \big )^{-1}
\mbox{     }.
\en
Note that since the sandpile will only be in a state of
incipient failure below some fixed depth, the stress tensor retains some
indeterminacy above that depth. Our Mohr analysis,  now local,  applies only at the incipent failure
depth
and does not determine the global stress state of the sand pile.

The criterion thus derived,
Eq.\ (\ref{failure_criterion}), is most stringent as
$z
\to
\infty$, in which case the dry sandpile result $\theta_c = \tan^{-1} k$ is
recovered. On the other hand, for a sandpile of fixed depth $D$, the
failure must
occur at most at depth
$D$.  Thus the critical angle will be the solution of the equation
\eq
k = \tan \theta_c(D) \big( 1+ \frac{s_A}{\rho g D \cos \theta_c(D)}
\big )^{-1} \mbox{   },
\en
giving an critical angle $\theta_c(D)$ that decreases monotonically
with
$D$. Thus finite humid sandpiles have a depth-dependent critical angle, unlike
dry sandpiles, which is a well-known result in soil mechanics \cite{class}. In
addition, we see that humid ({\em i.e.} cohesive) sandpiles fail at depth,
whereas from statics alone we were unable to determine the failure depth of dry
sandpiles. In the case of small adhesion stress $s_A/\rho g D \ll 1$, we
can write
\eq
\label{eq:answer}
\tan \theta_c \approx k + \frac{k s_A}{\rho g D} \sec \left[ \tan^{-1} (k)
\right]
\mbox{   }.
\en

\begin{figure}
\centerline{\psfig{figure=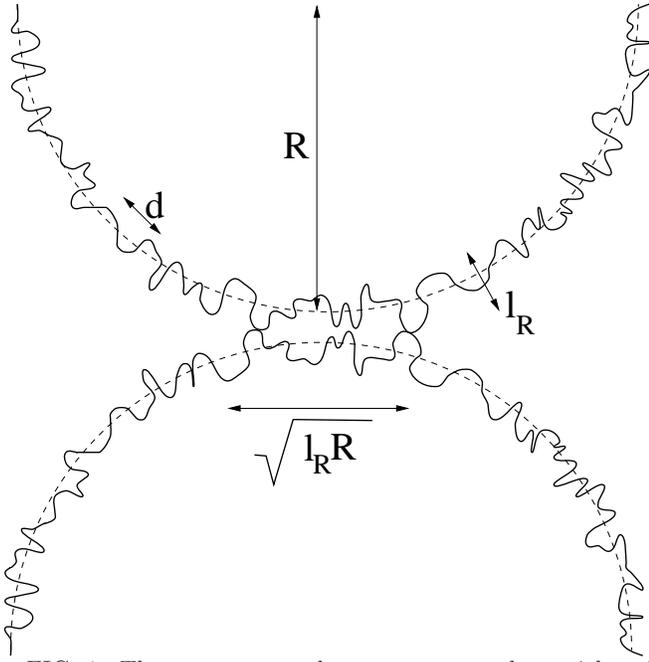,width=\hsize}}
\caption{The contact zone between two rough particles of radius $R$.  The
scale of
height deviations from the mean is $l_R$ and the height fluctuations are
correlated over a distance $d$.   The lateral size of the
contact zone in which the macroscopic curvature of the particles is not
apparent is
$\sim  \protect\sqrt{l_R R}$.}
\label{grain_pic}
\end{figure}

If, as we are assuming, the adhesion arises from capillary forces, we must
still
connect the adhesive stress to the amount of fluid present.
We suppose that the sand is composed of macroscopically spherical grains
(radius
$R$) whose surface roughness may be characterized as follows:  the spatial
correlation of fluctuations in local surface height saturates at height
$l_R$ at a
lateral distance $d$ that is much smaller than the particle radius, $d \ll
R$.   Since the particles are macroscopically spherical, we require that $ l_R
\ll R$ (see Fig.\ \ref{grain_pic}).

We can characterize the surface roughness of two particles in contact by
considering the function
$\delta(x)$ which gives the average distance between the two particles a
lateral distance $x$ from an asperity at which the two particles are in
contact. We write $\delta(x)$ in the form \cite{rough}
\eq
\label{delta}
\delta(x) = l_R f(x/d) \mbox{  },
\en
where $f(w)$ is a scaling function with the limits
\eq
f(w) \sim
\cases{ w^\chi & $w \rightarrow 0$, \cr
$1$       &
$w \rightarrow \infty$\mbox{  }. }
\en
The roughness exponent, $\chi$, of the surface satisfies $0 < \chi \le 1$.

Note that this form can only be valid for $x < \sqrt{l_R R}$. For larger values
of
$x$, the macroscopic curvature of the particles will determine the local
distance
between them (see Fig.\ \ref{grain_pic}).

For such particles, there are three
regimes for the capillary force exerted by a wetting fluid
as a function of $V$, the total amount of fluid present per particle contact.

\paragraph*{Asperity Regime--} For the smallest values of $V$, the
capillary force is
dominated by the accumulation of fluid around a single or a small number of
asperities at which two neighboring particles are in contact. This will hold
until the lateral extent of the fluid-filled region exceeds $d$, determining a
maximum contact fluid volume for this regime, $V_1 \sim l_R d^2$.

We can write the adhesion force $f_A$ as
\eq
\label{co-force}
f_A = \frac{\Gamma A}{r} \mbox{  },
\en
where $\Gamma$ is the surface tension of the fluid, $r$ is the radius of
curvature
of the meniscus of the fluid layer connecting the two grains near the
asperity and
$A
\sim V/\delta $ is the area of the contact patch. $-\Gamma/r$ is the pressure
reduction due to the capillary meniscus. Because $r$ will be approximately the
distance between the particles at the meniscus, we find
\eq
P \sim - \frac{\Gamma}{l_R}\left(\frac{V}{V_1}\right)^{- \frac{\chi}{2 +\chi}}
\mbox{  },
\en
and the adhesion force is
\eq
\label{regime1_f}
f_A \sim \frac{\Gamma \, V_1}{l_R^2} \left(\frac{V}{V_1}
\right)^{\frac{2 -\chi}{2 +\chi}} \mbox{, for  $V < V_1$,}
\en
where $V_1 = l_R {d}^2 $.

For a rough surface where $\chi = 1$,  Eq.~(\ref{regime1_f}) shows that the
force
depends on the cube root of the fluid volume. This is identical to the
dependence
of the cohesive force between a cone and plate on the volume of the liquid
bridge
connecting them \cite{ndame}.  It is to be expected that the cone-and-plate
model will reproduce the cohesive force near a single asperity.

\paragraph*{Roughness Regime--} For larger values of $V$, the fluid will
occupy a
statistically rough region, which is still small enough that the macroscopic
curvature of the particles plays no role--however, the fluid occupies more than
the area around a single asperity. This regime occurs for $V_1 < V < V_2$,
where
$V_2 = l_R^2 R$. The pressure is
\eq
P \sim - \frac{\Gamma}{l_R} \mbox{  },
\en
and the force will be
\eq
\label{regime2_f}
f_A \sim \frac{ \Gamma \, V}{l_R^2} \mbox{, for $ V_1 <V <
 V_2$} \mbox{   },
\en
In this roughness regime, the cohesive
force is linear in the volume of the added fluid, reproducing the linear
dependence found by Hornbaker {\em et al.}

\begin{figure}
\centerline{\psfig{figure=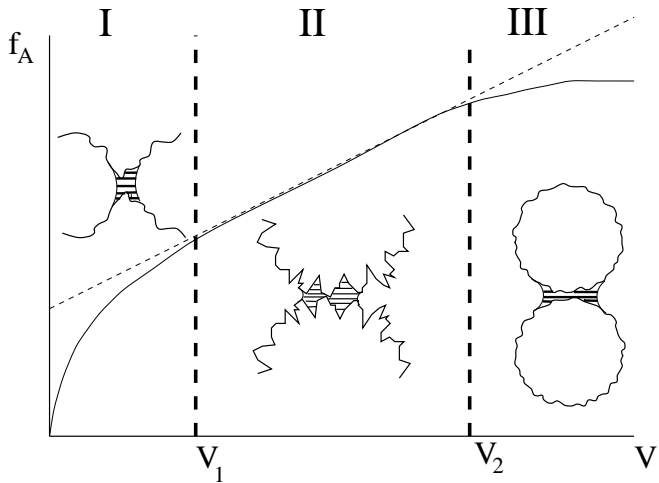,width=\hsize}}
\caption{The behavior of the adhesive force between two
rough, ``spherical" particles.  The three regimes of the force vs. volume of the
wetting layer are  I --  Asperity Regime,
II --  Roughness Regime, and III --  Spherical Regime.  The insets show the
extent
of the wetting region typical of each regime.}
\label{graph}
\end{figure}

\paragraph*{Spherical Regime--} When the lateral extent of the fluid contact
exceeds $d$, then the wetting region will be determined by the macroscopic
curvature of the particles, and the surface roughness will no longer play a
significant role. In this case the pressure is given by \cite{tabor}
\eq
P = -\frac{\Gamma}{\sqrt{V/2 \pi R}} \mbox{   },
\en
and the force by
\eq
\label{regime3_f}
f_A = 2 \pi \Gamma R \mbox{, for $ V >
 V_2$} \mbox{   },
\en
which is {\em independent} of the volume of the liquid bridge joining the two
grains.  Thus the linear increase of the cohesive force with fluid volume
saturates
for volumes $V > V_2 = l_R^2 R$ (see Fig.\ \ref{graph}).

If the fluid wets the surface of the particles, then in addition to the
fluid in
the contact region, there will also be a layer of fluid of thickness $t$ on the
surface of the particles. Typically this film will be no thicker than a few
monolayers; hence there are a complicated set of forces between
this film
and the surface. To simplify, we consider only the Van der Waals forces,
which for a thickness $t$ generate a ``disjoining pressure" $P_d$ given by
\cite{safran}
\eq
P_d = \frac{2H}{t^3} \mbox{  },
\en
where $H$ is the Hamaker constant ($H<0$ for a wetting fluid). To determine the
thickness of the wetting region, we must set this disjoining pressure equal
to the
pressure inside the contact regions. If the radius of curvature of the contact
meniscus is $r$, then
\eq
t =\left(- \frac{ 2 H r}{\Gamma}\right)^{\frac{1}{3}} \mbox{  }.
\en

In the asperity regime, $t$ will increase with the meniscus radius of curvature. In
the roughness regime, however, $r$ saturates at a value $\sim l_R$, and $t$
will be constant. In this regime any added fluid enters the contact region.
Finally, in the spherical regime, $t$ will again increase.

Now consider a system with a volume $V_l$ of liquid per particle. If we suppose
that the spheres are close-packed, then each sphere has 12 neighbors, so there
is an average of 6 contacts per sphere, each with a fluid volume of $V=V_l/6$.
The average number of contacts per unit area will be $(3 \phi_V/ \pi R^2)$,
where $\phi_V$ is the volume fraction of the particles. Thus for a
close-packed lattice, for which $\phi_V = \sqrt{2} \pi/6$, the adhesive stress
will be approximately
\eq
s_A \approx \frac{f_A}{\sqrt{2} R^2},
\en
and we can now substitute into Eq.\ (\ref{eq:answer}) to obtain the dependence
of the critical angle upon fluid volume. For a sandpile of fixed depth $D$,
this
is linear in $V$,
\eq
\label{main_ans}
\tan \theta_c  \approx k + \frac{ k \Gamma V }{\sqrt{2} \, l_R^2 R^2 \rho g D}
\sec\left[ \tan^{-1} (k) \right] \mbox{   },
\en
up to a saturation result determined by (see Fig.\ \ref{graph})
\eq
\tan \theta_c \approx k + \frac{ \sqrt{2} \,  \pi k \Gamma}{R \rho g D}
\sec \left[ \tan^{-1} (k) \right] \mbox{   }.
\label{eq:sat}
\en

Now let us consider the experiment of Hornbaker {\em et al.} They measured the
critical angle for a medium composed of radius $4
\times 10^{-2} \,
\cm$ polystyrene spheres with small amounts of added oil using the draining
crater method \cite{ndame}. They found a linear increase in the critical angle
measured as a function of the volume of added oil. They claimed that the
failure of their systems was at the surface, and concluded that they could
account for their results by assuming that 99.9\% of the fluid was outside of
the contact zones between the particles. Their particles had a surface
roughness
on the order of $1 \mu
\mbox{m}$.

We find that the increase in the critical angle is linear with the fluid
volume in the roughness regime, up to the saturation result Eq.\
(\ref{eq:sat}).
We expect the Hamaker constant for a wetting fluid to be negative, and of
the order
of magnitude of $H
\sim 10^{-20} \, \erg$, so no more than a few monolayers of fluid should be
present
on the particle surfaces. Furthermore, in the linear regime, all fluid
added to the
system will enter the particle contacts. Thus we disagree with the claim
that the
overwhelming majority of the fluid in this case will be outside the contact
zones.
Finally, we disagree with the interpretation of their experiment, according to
which surface failure is the most relevant failure mode-- the
failure plane in cohesive materials should be at depth.

We are grateful to P. Schiffer for providing us with copies of Ref.
\cite{ndame} before publication and to D. Erta{\c s} for many useful
discussions.

\end{document}